\documentclass[prd,aps,showpacs,preprintnumbers,amssymb]{revtex4}
\usepackage{axodraw}
\usepackage{color}
\usepackage{epsf}

\def\e3p{$\eta \rightarrow 3 \pi$}

\begin{document}
\title{%
\hfill{\normalsize\vbox{%
\hbox{}
 }}\\
{A nonperturbative method for QCD}}

\author{Renata Jora
$^{\it \bf a}$~\footnote[2]{Email:
 rjora@theory.nipne.ro}}

\affiliation{$^{\bf \it b}$ National Institute of Physics and Nuclear Engineering PO Box MG-6, Bucharest-Magurele, Romania}

\date{\today}

\begin{abstract}
Based  on specific properties of the partition function and of the quantum correlators we derive the exact form of the beta function in the background gauge field method for QCD with an arbitrary number of flavors.  The all order beta function we obtain through this method has only the first two orders coefficients different than zero and thus is equivalent to the 't Hooft scheme.
\end{abstract}
\maketitle

\section{Introduction}
 The running of the gauge coupling constant with the scale is generally computed in quantum field theories using a perturbative approach which consists in expansion in a small parameter. In the dimensional regularization scheme beta function for QED is known at the fourth order whereas that for QCD at the fifth one \cite{Vladimirov}-\cite{Baikov}. It is known that the first two order coefficients are renormalization scheme independent whereas the next ones depend on the specific renormalization procedure. In \cite{Hooft1}, \cite{Hooft2} 't Hooft introduced a procedure in which the beta function stops at two loops.

 In \cite{Jora1}we used  a semi perturbative technique to compute the beta function for QED to obtain that the beta function stops at two loops.  This method was further simplified and improved in \cite{Jora} where the exact form of the beta function for the Yang Mills theory has been determined from properties of the partition function and various correlators  and by the use of the LSZ theorem. In the present work we extend our method to the more intricate case of the beta function of QCD with an arbitrary number of flavors. Whereas this case require more work the principles settled in \cite{Jora} remain unaltered. We determine that as in \cite{Jora1}, \cite{Jora} the all orders beta function stops at the two first orders coefficients. Note that this result is obtained without using any Feynman diagram or expansion in a small parameter.

\section{Partition function for an $SU(N)$ theory with $N_f$ flavors}
We start with the gauge fixed Lagrangian for an $SU(N)$ gauge theory with $N_f$ fermions in the fundamental representation:
\begin{eqnarray}
{\cal L}=-\frac{1}{4}(F^a_{\mu\nu})^2+\bar{c}^a(-\partial^{\mu}\partial_{\mu}-g f^{abc}\partial^{\mu}A^b_{\mu})c^c+\sum_f\bar{\Psi}(i\gamma^{\mu}D_{\mu}-m_f)\Psi,
\label{rez5467}
\end{eqnarray}
where,
\begin{eqnarray}
F^a_{\mu\nu}=\partial_{\mu}A^a_{\nu}-\partial_{\nu}A^a_{\mu}+gf^{abc}A^b_{\mu}A^c_{\nu},
\label{ten45678}
\end{eqnarray}
and,
\begin{eqnarray}
D_{\mu}=\partial_{\mu}-igA^a_{\mu}t^a.
\label{cov554}
\end{eqnarray}
Here $t^a$ is the generator of $SU(N)$ in the fundamental representation and for simplicity we shall consider $m_f=0$.
We shall work in the Feynman gauge ($\xi=1$). One can express all the fields in the Fourier space:
\begin{eqnarray}
&&A^a_{\mu}(x)=\frac{1}{V}\sum_n\exp[-i k_n x]A^a_{\mu}(k_n)
\nonumber\\
&&\Psi(x)=\frac{1}{V}\sum_m\exp[-i k_m x]\Psi(k_m)
\nonumber\\
&&c^b(x)=\frac{1}{V}\sum_p\exp[-i k_p x]c^b(k_p).
\label{fourier456}
\end{eqnarray}
Then the Lagrangian takes the form:
\begin{eqnarray}
&&\int d^4x {\cal L}=-\frac{1}{2}\frac{1}{V} \sum_n k_n^2A^{a\nu}(k_n)A^a_{\nu}(-k_n)+\frac{1}{V}\sum_n k_n^2 \bar{c}^a(k_n)c^a(-k_n)+
\nonumber\\
&&+\frac{i}{V^2}g\sum_{n,m} k_n^{\mu}A^a_{\nu}(k_n)f^{abc}A^b_{\mu}(k_m)A^{c\nu}(-k_n-k_m)-
\nonumber\\
&&-\frac{1}{V^3}g^2f^{abc}f^{ade}\sum_{n,m,p}A^{b\mu}(k_n)A^{c\nu}(k_m)A^d_{\mu}(k_p)A^e_{\nu}(-k_n-k_m-k_p)-
\nonumber\\
&&-\frac{i}{V^2}\sum_{n,m}k_n^{\mu}\bar{c}^a(k_n)gf^{abc}A^b_{\mu}(k_m)c^c(-k_n-k_m)+
\nonumber\\
&&\frac{1}{V}\sum_f\sum_n\bar{\Psi}_f(k_n)\gamma^{\mu}k_{\mu n}\Psi_f(k_n)+\frac{1}{V^2}g\sum_f\sum_{n,m}\bar{\Psi}_f(k_n)\gamma^{\mu}A^a_{\mu}(k_n-k_m)t^a\Psi_f(k_m).
\label{four65788}
\end{eqnarray}

The zero current partition function has the form:
\begin{eqnarray}
Z_0=\int \prod_i\prod_j\prod_m \prod_{fl}\prod_n\prod_p d A^a_{\mu}(k_i) d \bar{c}^b(k_j)d c^d(k_m)d\bar{\Psi}_{fl}(k_n) d\Psi_{fl}(k_p)\exp[i\int d^4x {\cal L}],
\label{part456}
\end{eqnarray}
where the exponent is considered in the Fourier space.

It is useful at this stage to settle some of the properties of $Z_0$. It is known that $Z_0$ apart from a factor in front is given by the exponential of the sum of all disconnected diagrams:
\begin{eqnarray}
Z_0={\rm factor}\times \exp[\sum_i V_i]
\label{disc4567}
\end{eqnarray}
where $V_i$ is a typical disconnected diagram. Since the calculation is done in the absence of external sources all $V_i$ diagrams are closed and contain summations over momenta (that appear in propagators or vertices) and thus do not depend at all on any momenta. The factor in front is a product obtained from integrating the gaussian integrals corresponding to the kinetic terms.
The final result has thus the expression:
\begin{eqnarray}
Z_0= {\rm const} \prod_i (k_i^2)^{N^2-1}\prod_j (k_j^2)^{-d/2(N^2-1)}\prod_p (\gamma^{\mu}p_{\mu}-m)^{N_fN}\exp[\sum_i V_i]
\label{part4567}
\end{eqnarray}
where N is coming from the Yang Mills group $SU(N)$ and the first factor corresponds to the ghosts, the second to the gluon fields and the  third to the fermion fields.

We shall apply the same procedure as in \cite{Jora} to determine whole properties of the partition function and of the fields.
First we consider the partition function in Eq. (\ref{part456}) and introduce in the integrand the quantity $\frac{d A^a_{\nu}(k)}{d A^a_{\nu}(k)}$ to obtain:

\begin{eqnarray}
&&Z_0=\int \prod_{fl} \prod_i \prod_j \prod_n\prod_p\prod_m d A^a_{\mu}(k_i) d \bar{c}^b(k_j) d c^d (k_m) d\bar{\Psi}_{fl}(k_n) d\Psi_{fl}(k_p)\exp[i \int d^4x{\cal L}]=
\nonumber\\
&&=\int \prod_{fl} \prod_i \prod_j \prod_m \prod_n\prod_p d A^a_{\mu}(k_i) d \bar{c}^b(k_j) d c^d (k_m) d\bar{\Psi}_{fl}(k_n) d\Psi_{fl}(k_p)\frac{ d A^a_{\nu}(k)}{d A^a_{\nu}(k)}\exp[i \int d^4x{\cal L}]=
\nonumber\\
&&= \int \prod_{fl} \prod_i \prod_j \prod_m \prod_n\prod_p d A^a_{\mu}(k_i) d \bar{c}^b(k_j) d c^d (k_m) d\bar{\Psi}_{fl}(k_n) d\Psi_{fl}(k_p)\frac{d}{d A^a_{\nu}(k)}[ A^a_{\nu}(k)\exp[i \int d^4x{\cal L}]]-
\nonumber\\
&&-\int \prod_{fl} \prod_i \prod_j \prod_m \prod_n\prod_pd A^a_{\mu}(k_i) d \bar{c}^b(k_j) d c^d (k_m) d\bar{\Psi}_{fl}(k_n) d\Psi_{fl}(k_p)A^a_{\nu}(k)\frac{d}{d A^a_{\nu}(k)}\exp[i \int d^4x{\cal L}],
\label{rez54678}
\end{eqnarray}
where $\prod_{fl}$ contains separate products over flavors and colors.

The first term on the right side of the Eq. (\ref{rez54678}),
\begin{eqnarray}
&&\int \prod_{fl}\prod_i \prod_j \prod_m \prod_n\prod_p dA^d_{\mu}(k_i) d \bar{c}^b(k_j) d c^d (k_m) d\bar{\Psi}_f(k_n) d\Psi(k_p)A^a_{\nu}(k)\exp[i \int d^4x{\cal L}]_{A^a_{\nu}(k)=+\infty}-
\nonumber\\
&&\int \prod_{fl}\prod_i \prod_j \prod_m \prod_n\prod_p d A^d_{\mu}(k_i) d \bar{c}^b(k_j) d c^d (k_m)d\bar{\Psi}_f(k_n) d\Psi(k_p) A^a_{\nu}(k)\exp[i \int d^4x{\cal L}]_{A^a_{\nu}(k)=-\infty},
\label{expr7689}
\end{eqnarray}
 is zero since the $\epsilon$ term in the kinetic term will lead to an exponential that goes to zero (see \cite{Jora} for more detailed explanation). Here the product satisfy the constraint: $A^d_{\mu}(k_i)\neq A^a_{\mu}(k)$).

  The second contribution leads to the result:
\begin{eqnarray}
&&Z_0=\int \prod_{fl}\prod_i \prod_j \prod_m \prod_n \prod_p d A^a_{\mu}(k_i) d \bar{c}^b(k_j) d c^d (k_m)d\bar{\Psi}_{fl}(k_n) d\Psi_{fl}(k_p)(-i)[-\frac{k^2}{V}A^{a\nu}(k)A^a_{\nu}(-k)+
\nonumber\\
&&\frac{3i}{V^2}gk^{\mu}\sum_pf^{abc}A^a_{\nu}(k)A^b_{\mu}(p)A^{c\nu}(-k-p) -\frac{i}{V^2}g\sum_pp^{\nu} \bar{c}^b(p)f^{bac}A^{a}_{\nu}(k)c^c(-p-k)-
\nonumber\\
&&-\frac{1}{V^3}g^2f^{bac}f^{bde}\sum_{p,q}A^a_{\nu}(k)A^c_{\mu}(p)A^{d\nu}(q)A^{e\mu}(-p-k-q)+
\nonumber\\
&&g\frac{1}{V^2} \sum_p\bar{\Psi}(p)\gamma^{\mu}t^aA^a_{\mu}(k)\Psi(-p-k)]\times\exp[i\int d^4x {\cal L}].
\label{one65789}
\end{eqnarray}

 We apply the same procedure to the partition function but this time introduce in the integrand the quantity $\frac{d \bar{\Psi}_{f_1}^r}{d \bar{\Psi}_{f_1}^r}$, where $f$ is a flavor index and $r$ is a color one. This yields:
 \begin{eqnarray}
&&Z_0=-i\int \prod_{fl}\prod_i \prod_j\prod_m \prod_n \prod_p d A^a_{\mu}(k_i) d \bar{c}^b(k_j) d c^d (k_m)d\bar{\Psi}_{fl}(k_n) d\Psi_{fl}(k_p)\times
\nonumber\\
&&[\frac{1}{V}\bar{\Psi}_f^r\gamma^{\mu}k_{\mu}\Psi_f^r+\frac{1}{v^2}\sum_p\bar{\Psi}^r_f(k)\gamma^{\mu}p_{\mu}t^a_{rj}A^a_{\mu}(-p+k)\Psi_f^j(p)]\exp[i\int d^4x {\cal L}].
\label{fermrel8767}
\end{eqnarray}
Here we used the fact:
\begin{eqnarray}
 -i\int \prod_{fl}\prod_i \prod_j \prod_m \prod_n \prod_p d A^a_{\mu}(k_i) d \bar{c}^b(k_j) d c^d (k_m)d\bar{\Psi}_{fl}(k_n) d\Psi_{fl}(k_p)\bar{\Psi}^r_{f_1}(k)\exp[i\int d^4x {\cal L}]|_{\bar{\Psi}^i_f(k)\rightarrow\pm\infty}=0
 \label{cond556}
 \end{eqnarray}
 since the spinors fields anticommute and there is no pairing for $\Psi^r_{f_1}(k)$ (there is no integration over $\bar{\Psi}^r_{f_1}(k)$) and thus the result is zero.

A similar procedure applied to the ghost field $c^c(k)$ leads to:
\begin{eqnarray}
&&Z_0=-i\int \prod_{fl}\prod_i \prod_j \prod_m \prod_n \prod_pd A^a_{\mu}(k_i) d \bar{c}^b(k_j) d c^d (k_m)d\bar{\Psi}_{fl}(k_n) d\Psi_{fl}(k_p)\times
\nonumber\\
&&[\frac{1}{V}k^2\bar{c}^c(k)k^2c^c(k)-\frac{i}{V^2}g\sum_p\bar{c}^a(p)p^{\mu}f^{abc}c^c(k)A^b_{\mu}(p-k)]\exp[i\int d^4x {\cal L}].
\label{ghost677}
\end{eqnarray}

Next we apply the operator $k^{\mu}\frac{d}{d k^{\mu}}$ to the Eq. (\ref{part456}) to obtain:
\begin{eqnarray}
&&k^{\mu}\frac{d Z_0}{d k^{\mu}}=\int \prod_{fl}\prod_i \prod_j \prod_m \prod_n \prod_p d A^a_{\mu}(k_i) d \bar{c}^b(k_j) d c^d (k_m)d\bar{\Psi}_{fl}(k_n) d\Psi_{fl}(k_p)\times
\nonumber\\
&&i[-\frac{1}{V}k^2A^{a\nu}(k)A^a_{\nu}(-k)+\frac{2}{V}k^2\bar{c}^a(k)c^a(-k)+\frac{i}{V^2}k^{\mu}\sum_{p}A^a_{\nu}(k)f^{abc}gA^b_{\mu}(p)A^{c\nu}(-p-k)-
\nonumber\\
&&-\frac{i}{V^2}\sum_{p}k^{\mu}\bar{c}^a(k)gf^{abc}A^b_{\mu}(p)c^c(-p-k)+\frac{1}{V}\bar{\Psi}(k)\gamma^{\mu}k_{\mu}\Psi(k)]\times\exp[i\int d^4x {\cal L}],
\label{sec4355}
\end{eqnarray}
where from Eq. (\ref{part4567}) we calculate:
\begin{eqnarray}
k^{\mu}\frac{d Z_0} {\partial k^{\mu}}=\left[N_f N-2(N^2-1)[\frac{d}{2}-1]\right]Z_0.
\label{some324}
\end{eqnarray}

\section{Renormalization}

In this section we shall consider all the results in section II from the perspective of renormalization. Thus the renormalized Lagrangian is:
\begin{eqnarray}
&&\int d^4 x{\cal L}_r=-\frac{1}{2}\frac{1}{V}Z_3 \sum_n k_n^2A^{a\nu}(k_n)A^a_{\nu}(-k_n)+\frac{1}{V}Z_1\sum_n k_n^2 \bar{c}^a(k_n)c^a(-k_n)+
\nonumber\\
&&+\frac{i}{V^2}Z_{3g}g\sum_{n,m} k_n^{\mu}A^a_{\nu}(k_n)f^{abc}A^b_{\mu}(k_m)A^{c\nu}(-k_n-k_m)-
\nonumber\\
&&-\frac{1}{V^3}Z_{4g}g^2f^{abc}f^{ade}\sum_{n,m,p}A^{b\mu}(k_n)A^{c\nu}(k_m)A^d_{\mu}(k_p)A^e_{\nu}(-k_n-k_m-k_p)-
\nonumber\\
&&-\frac{i}{V^2}Z_1'\sum_{n,m}k_n^{\mu}\bar{c}^a(k_n)gf^{abc}A^b_{\mu}(k_m)c^c(-k_n-k_m)+
\nonumber\\
&&\frac{1}{V}Z_2\sum_f\sum_n\bar{\Psi}_f(k_n)\gamma^{\mu}k_{\mu n}\Psi_f(k_n)+\frac{1}{V^2}g Z_2'\sum_f\sum_{n,m}\bar{\Psi}_f(k_n)\gamma^{\mu}A^a_{\mu}t^a\Psi(k_m)_f.
\label{four6522788}
\end{eqnarray}
where for simplicity we drop the index $r$ from the renormalized fields.

Then Eq. (\ref{one65789}) will become:
\begin{eqnarray}
&&Z_0=\int \prod_{fl}\prod_i \prod_j \prod_m \prod_n \prod_p d A^a_{\mu}(k_i) d \bar{c}^b(k_j) d c^d (k_m)d\bar{\Psi}_{fl}(k_n) d\Psi_{fl}(k_p)(-i)[-Z_3\frac{k^2}{V}A^{a\nu}(k)A^a_{\nu}(-k)+
\nonumber\\
&&Z_{3g}\frac{3i}{V^2}gk^{\mu}\sum_pf^{abc}A^a_{\nu}(k)A^b_{\mu}(p)A^{c\nu}(-k-p) -\frac{i}{V^2}Z_1'g\sum_pp^{\nu} \bar{c}^b(p)f^{bac}A^{a}_{\nu}(k)c^c(-p-k)-
\nonumber\\
&&-\frac{1}{V^3}g^2Z_{4g}f^{bac}f^{bde}\sum_{p,q}A^a_{\nu}(k)A^c_{\mu}(p)A^{d\nu}(q)A^{e\mu}(-p-k-q)+
\nonumber\\
&&g\frac{1}{V^2}Z_2'g \sum_p\bar{\Psi}(p)\gamma^{\mu}t^aA^a_{\mu}(k)\Psi(-p-k)]\times\exp[i\int d^4x {\cal L}].
\label{one657891}
\end{eqnarray}

Eq. (\ref{fermrel8767}) will transform to,
 \begin{eqnarray}
&&Z_0=-i\int \prod_{fl}\prod_i \prod_j \prod_m \prod_n \prod_p d A^a_{\mu}(k_i) d \bar{c}^b(k_j) d c^d (k_m)d\bar{\Psi}_{fl}(k_n) d\Psi_{fl}(k_p)\times
\nonumber\\
&&[\frac{1}{V}Z_2\bar{\Psi}_f^r\gamma^{\mu}k_{\mu}\Psi_f^r+\frac{1}{V^2}Z_2'g\sum_p\bar{\Psi}^r_f(k)\gamma^{\mu}p_{\mu}t^a_{rj}A^a_{\mu}(-p+k)\Psi_f^j(p)]\exp[i\int d^4x {\cal L}].
\label{fermrel87672}
\end{eqnarray}

whereas Eq. (\ref{ghost677}) yields:
\begin{eqnarray}
&&Z_0=-i\int \prod_{fl}\prod_i \prod_j \prod_m \prod_n \prod_p d A^a_{\mu}(k_i) d \bar{c}^b(k_j) d c^d (k_m)d\bar{\Psi}_{fl}(k_n) d\Psi_{fl}(k_p)\times
\nonumber\\
&&[\frac{1}{V}Z_1k^2\bar{c}^c(k)k^2c^c(k)-\frac{i}{V^2}g Z_1'\sum_p\bar{c}^a(p)p^{\mu}f^{abc}c^c(k)A^b_{\mu}(p-k)]\exp[i\int d^4x {\cal L}].
\label{ghost6778}
\end{eqnarray}

Finally Eq. (\ref{sec4355}) will lead to:
\begin{eqnarray}
&&k^{\mu}\frac{d Z_0}{d k^{\mu}}=\int \prod_{fl}\prod_i \prod_j \prod_m \prod_n \prod_p d A^a_{\mu}(k_i) d \bar{c}^b(k_j) d c^d (k_m)d\bar{\Psi}_{fl}(k_n) d\Psi_{fl}(k_p)\times
\nonumber\\
&&i[-\frac{1}{V}Z_3k^2A^{a\nu}(k)A^a_{\nu}(-k)+Z_1\frac{2}{V}k^2\bar{c}^a(k)c^a(-k)+\frac{i}{V^2}Z_{3g}k^{\mu}\sum_{p}A^a_{\nu}(k)f^{abc}gA^b_{\mu}(p)A^{c\nu}(-p-k)-
\nonumber\\
&&-\frac{i}{V^2}Z_1'\sum_{p}k^{\mu}\bar{c}^a(k)gf^{abc}A^b_{\mu}(p)c^c(-p-k)+\frac{1}{V}Z_2\bar{\Psi}(k)\gamma^{\mu}k_{\mu}\Psi(k)]\times\exp[i\int d^4x {\cal L}].
\label{sec435532}
\end{eqnarray}

\section{Relations among the renormalization constants}

In the path integral formalism the two point gluon function has the expression:
\begin{eqnarray}
&&\langle\Omega | T[A^a_{\mu}(x_1)A^b_{\nu}(x_2)] |\Omega\rangle=
\nonumber\\
&&\lim_{T\rightarrow \infty(1-i\epsilon)}\frac{\int d A^c_{\rho} d\bar{c}^d d c^e d\bar{\Psi}d\Psi A^a_{\mu}(x_1)A^b_{\nu}(x_2)\exp[i\int d^4x {\cal L}]}
{\int d A^c_{\rho}d\bar{c}^d d c^e d\bar{\Psi}d\Psi\exp[i\int d^4x {\cal L}]}.
\label{equibrel65789}
\end{eqnarray}

We apply the LSZ reduction formula in the path integral formalism and in the Fourier space:
\begin{eqnarray}
&&\langle \Omega|T[A^a_{\mu}(p_1)...A^d_{\nu}(p_m)A^b_{\rho}(k_1)...A^e_{\sigma}(k_n)]|\Omega\rangle\sim
\nonumber\\
&&\sim_{p_i^0(k_j^0)\rightarrow E_{\vec{p}_i}(E_{\vec{k}_j})}{\rm polarization\,factor}\times{\rm const}\times\langle \vec{p}_1...\vec{p}_m|S|\vec{k}_1...\vec{k}_n\rangle
\left(\prod_{i=1}^m\frac{i Z_3^{1/2}}{p_i^2+i\epsilon}\right) \left(\prod_{j=1}^n\frac{i Z_3^{1/2}}{k_j^2+i\epsilon}\right).
\label{LSZred546}
\end{eqnarray}

In \cite{Jora} we illustrate in detail how  we apply this formula to the gauge and ghost terms in the relations in Eqs. (\ref{one657891}), (\ref{fermrel87672}), (\ref{ghost6778}) and (\ref{sec435532}).

LSZ formula is more intricate and complicated for fermions as it can be seen from the following equation for a process with two initial and two final fermions \cite{Srednicki}:
\begin{eqnarray}
&&_{out}{\langle} f|i \rangle_{in}=\langle f|S|i\rangle\approx
\int d^4 x_1 d^4 x_2 d^4 y_1 d^4 y_2 \exp[-i k_1 y_1][\bar{u}_{s_1^{\prime}}(k_1)(-i\gamma^{\mu}{\partial}_{\mu y_1}+m)]_{\beta_1}\times
\nonumber\\
&&\exp[-i k_2 y_2][\bar{u}_{s_2^{\prime}}(k_2)(-i\gamma^{\mu}{\partial}_{\mu y_2}+m)]_{\beta_2}\times
\nonumber\\
&&\langle 0 |T\Psi_{\beta_2}(y_2)\Psi_{\beta_1}(y_1)\bar{\Psi}_{\alpha_1}(x_1)\bar{\Psi}_{\alpha_2}(x_2)| 0\rangle\times
\nonumber\\
&&[(i\gamma^{\mu}\overleftarrow{\partial}_{\mu x_1}+m)u_{s_1}(p_1)]_{\alpha_1}]\exp[ip_1x_1]\times
\nonumber\\
&&[(i\gamma^{\mu}\overleftarrow{\partial}_{\mu x_2}+m)u_{s_2}(p_2)]_{\alpha_2}]\exp[ip_2x_2].
\label{fermions434}
\end{eqnarray}

Here $\alpha_1$, $\alpha_2$, $\beta_1$ and $\beta_2$ are spinor indices and all momenta are on shell.

This formula is too intricate to be easily applicable to our calculations. If $a_s^{\dagger}(\vec{p}$  and $b_s^{\dagger}(\vec{p})$ are the operators that create a one particle state with charge 1 respectively -1 one can write:
\begin{eqnarray}
&&a_s^{\dagger}(\vec{p})_{in} \rightarrow i \int d^4 x \bar{\Psi}(x)(i\gamma^{\mu}\overleftarrow{\partial}_{\mu}+m)u_s(\vec{p})\exp[i p x]
\nonumber\\
&&a_s(\vec{p})_{out} \rightarrow i \int d^4 x \exp[-i p x]\bar{u}_s(\vec{p}))(-i\gamma^{\mu}\partial_{\mu}+m)\Psi(x)
\nonumber\\
&&b_s^{\dagger}(\vec{p})_{in} \rightarrow i \int d^4 x\exp[i p x]\bar{v}_s(\vec{p}))(-i\gamma^{\mu}\partial_{\mu}+m)\Psi(x)
\nonumber\\
&&b_s(\vec{p})_{out} \rightarrow i \int d^4 x \bar{\Psi}(x)(i\gamma^{\mu}\overleftarrow{\partial}_{\mu}+m)v_s(\vec{p})\exp[-i p x].
\label{rez44343}
\end{eqnarray}
Let us rewrite the first equation in (\ref{rez44343}) in the Fourier:
\begin{eqnarray}
&&a_s^{\dagger}(\vec{p})=i \int d^4 x\int \frac{d^4 k}{(2\pi)^4}\bar{\Psi}(k)\exp[i k x](i\gamma^{\mu}\overleftarrow{\partial}_{\mu}+m)u_s(\vec{p}) \exp[ i p x]=
\nonumber\\
&&i \int d^4 x \int \frac{d^4k}{(2\pi)^4}\exp[i k x]\bar{\Psi}(k)(-\gamma^{\mu}k_{\mu}+m)u_s(\vec{p}) \exp[i p x]=
\nonumber\\
&&i\bar{\Psi}(p)(\gamma^{\mu}p_{\mu}+m)u_s(\vec{p}),
\label{rel77678}
\end{eqnarray}
The above formula is still useless as we need to express $\bar{\Psi}(\vec{p})$ in terms of the other quantities. In order to solve that we consider the sum:
\begin{eqnarray}
\sum_s a_s^{\dagger}(\vec{p})_{in}\bar{u}_s(\vec{p})=\sum_s i\bar{\Psi}(p)(\gamma^{\mu}p_{\mu}+m)u_s(\vec{p})\bar{u}_s(\vec{p}).
\label{rez221345}
\end{eqnarray}

Knowing that the following formula holds,
\begin{eqnarray}
\sum_s u_s(\vec{p})\bar{u}_s(\vec{p})=(-\gamma^{\mu}p_{\mu}+m),
\label{form77688}
\end{eqnarray}
we obtain:
\begin{eqnarray}
\sum_s a_s^{\dagger}(\vec{p})_{in}\bar{u}_s(\vec{p})= -i\bar{\Psi}(p)(p^2-m^2).
\label{rez443566}
\end{eqnarray}
From Eq. (\ref{rez443566}) we extract:
\begin{eqnarray}
\bar{\Psi}(p)=i \sum_sa_s^{\dagger}(\vec{p})_{in}\bar{u}_s(\vec{p})\frac{1}{p^2-m^2}.
\label{def3455}
\end{eqnarray}
As stated in section II we shall take $m_f=0$ in all subsequent calculations.

Now we shall apply all these findings to  Eq. (\ref{one657891}). The results from applying the LSZ reduction formula for the gauge fields and ghosts are detailed in \cite{Jora}. Here we
shall consider only the fermion fields. First we divide Eq. (\ref{one657891}) by $Z_0$ which yields:
\begin{eqnarray}
&&1={\rm terms\, that\, do \, not \,involve\, fermions}+
\frac{1}{Z_0}\int \prod_{fl}\prod_i \prod_j \prod_m \prod_n \prod_p d A^a_{\mu}(k_i) d \bar{c}^b(k_j) d c^d(k_m) d \bar{\Psi}_{fl}(k_n) d \Psi_{fl}( k_p)\times
\nonumber\\
&&(-i)g\frac{1}{V^2}Z_2' \sum_p \bar{\Psi}(p)\gamma^{\mu}t^a_rA^a_{\mu}\Psi(p-k) \exp[i \int d^4 x {\cal L}].
\label{res442324}
\end{eqnarray}
But the last term in Eq.(\ref{res442324}) is just :
\begin{eqnarray}
(-i)g\frac{1}{V^2}Z_2'\sum_p\langle \Omega|T[\bar{\Psi}(p)\gamma^{\mu}t^a_rA^a_{\mu}(k)\Psi(p-k)|\Omega\rangle.
\label{rez3989}
\end{eqnarray}
Then by applying Eqs. (\ref{fermions434}) and  (\ref{def3455}) to Eq. (\ref{rez3989}) one obtains for Eq. (\ref{res442324}):
\begin{eqnarray}
&&1={\rm terms\, that\, do \, not\, involve\, fermions}+
(-i)g\frac{1}{V^2}Z_2'\sum_p \frac{1}{p^2(p+k)^2k^2}\times {\rm const}
\nonumber\\
&&\sum_{s,s'}(t^a_r)_{ij}\bar{u}_s(\vec{p})\langle (\vec{p},s)_i;\vec{k}\epsilon_{k\mu}^a|\gamma^{\mu}S|((p+k),s')_j\rangle u_{s'}(\overrightarrow{p+k}).
\label{rezsd4356}
\end{eqnarray}
Note that in the above equation the term in brackets actually contains the vertex function $V^{a}_{ij}$ which is known by the renormalization conditions. Then one can write:
\begin{eqnarray}
\bar{u}(\vec{p})V^{a\mu}_{ji}\gamma_{\mu}(t^a_r)_{ij}u_s(\vec{p+k})\approx \frac{N^2-1}{2}gp^2
\label{res54677}
\end{eqnarray}
where we used the fact that for an on shell fermion $p^2=m^2$ and also as defined in the present work the vertex function ($V^{a\mu}_{ji}=gt^a_{ji}p^{\mu}$)contains already a compression between two fermion states.  Also note that the factor $\sum_p \frac{p^2}{k^2p^2(k+p)^2}$  in the limit of on shell states leads to a constant.

Since similar procedure (see also \cite{Jora}) applies to all the fields and interaction that appear in Eqs. (\ref{one657891}), (\ref{fermrel87672}), (\ref{ghost6778}) and  (\ref{sec435532}) these relations will become:
\begin{eqnarray}
&&1=aZ_3+b Z_{3g}g^2+c Z_1'g^2+d Z_{4g}g^4+g Z_2' g^2
\nonumber\\
&&s_1=s_2Z_2+ s_3 Z_2' g^2
\nonumber\\
&&r_1=r_2Z_1+r_3Z_1' g^2
\nonumber\\
&&x=yZ_3+z Z_1+ q Z_2+u Z_{3g}g^2+w Z_1'g^2,
\label{finalreljki7}
\end{eqnarray}
where we absorbed all the constants in front of the terms in the new coefficients $a$, $b$, $c$, $d$, $e$, $x$, $y$, $z$, $q$, $u$, $v$, $w$, $r_1$, $r_2$, $r_3$, $s_1$, $s_2$, $s_3$.

\section{Beta function and discussion}

In general in the dimensional regularization scheme similar relations exist also in other schemes) the renormalization constants satisfy the Slanov Taylor identities:
\begin{eqnarray}
g_0^2=\frac{Z_{3g}^2}{Z_3^3}g^2\mu^{\epsilon}=\frac{Z_{4g}}{Z_3^2}g^2\mu^{\epsilon}=
\frac{Z_1^{\prime 2}}{Z_1^2Z_3}g^2\mu^{\epsilon}=\frac{Z_2^{\prime 2}}{Z_2^2Z_3}g^2\mu^{\epsilon},
\label{sl7768}
\end{eqnarray}
where $d=4-\epsilon$ and $\mu$ is a parameter with dimension of mass.
In the background gauge field method there is a great simplification given by the relations:
\begin{eqnarray}
&&Z_1=Z_1'
\nonumber\\
&&Z_2=Z_2'
\nonumber\\
&&Z_3=Z_{3g}=Z_{4g}.
\label{bck6657}
\end{eqnarray}

Then one can write the four relations in Eq. (\ref{finalreljki7}) in a more compact form:
\begin{eqnarray}
&&1=(f_1+f_2g^2+f_3g^4)Z_3+f_4Z_1g^2+f_5Z_2g^2
\nonumber\\
&&1=(t_1+t_2g^2)Z_2
\nonumber\\
&&1=(c_1+c_2g^2)Z_1
\nonumber\\
&&1=(h_1+h_2g^2)Z_3+Z_1(h_3+h_4g^2)+h_5Z_2
\label{finalrel665466}
\end{eqnarray}

From the last three equations in Eq. (\ref{finalrel665466}) we determine:
\begin{eqnarray}
Z_3=\frac{(c_1t_1-h_3t_1-h_5c_1)+(c_2t_1+c_1t_2-h_3t_2-h_4t_1-h_5c_2)g^2+(c_2t_2-h_4t_2)g^4}{(h_1+h_2g^2)(c_1+c_2g^2)(t_1+t_2g^2)}.
\label{zres4434}
\end{eqnarray}

From the first three equations in Eq. (\ref{finalrel665466}) we compute:
\begin{eqnarray}
Z_3=\frac{c_1t_1+(c_2t_1+c_1t_2-f_4t_1-f_5c_1)g^2+(c_2t_2-f_4t_2-f_5c_2)g^4}{(f_1+f_2g^2+f_3g^4)(c_1+c_2g^2)(t_1+t_2g^2)}.
\label{seconds4356}
\end{eqnarray}
We shall use this last equation as a constraint. Matching the order of the coefficients with those in Eq. (\ref{zres4434}) we obtain the condition:
\begin{eqnarray}
(c_2t_2-h_4t_2)f_3=0
\label{constr5546}
\end{eqnarray}
from which we deduce $c_2=h_4$ since none of the coefficients are allowed to be zero.

There are some coefficients in Eq. (\ref{finalrel665466})  that can be determined directly form the preceding defining equations. These are those associated with the terms involving the gluon, fermion or ghost two point functions. Thus from Eq. (\ref{sec435532}) and the subsequent versions of it (noting that $h_1$, $h_3$ and $h_5$ are associated with the two point function for gluon, ghost and fermion respectively) one can compute by simple gaussian integration:
\begin{eqnarray}
&&h_1=\frac{-4(N^2-1)}{N N_f-2(N^2-1)}
\nonumber\\
&&h_3=\frac{2(N^2-1)}{N N_f-2(N^2-1)}
\nonumber\\
&&h_5=\frac{N_f N}{N N_f-2(N^2-1)}.
\label{first546}
\end{eqnarray}
Furthermore Eq. (\ref{first546}) leads to the following useful recurrence relation:
\begin{eqnarray}
1-h_3-h_5=h_1.
\label{rec4355}
\end{eqnarray}
Similarly from Eqs. (\ref{fermrel8767}) and (\ref{ghost6778}) one can determine $c_1=t_1=1$ for the same reasons.  By substituting in Eq.(\ref{zres4434}) the correct values for $h_1$, $h_3$, $h_5$, $c_1$, $t_1$ the expression for $Z_3$ becomes:
\begin{eqnarray}
Z_3=\frac{1+u_1g^2}{1+v_1g^2+v_2g^4+v_3g^6},
\label{finares4355}
\end{eqnarray}
where,
\begin{eqnarray}
&&u_1=\frac{t_2}{h_1}-\frac{h_3t_2}{h_1}-\frac{h_5c_2}{h_1}
\nonumber\\
&&v_1=c_2+t_2+\frac{h_2}{h_1}
\nonumber\\
&&v_2=\frac{h_2t_2}{h_1}+c_2t_2+\frac{h_2c_2}{h_1}
\nonumber\\
&&v_3=\frac{h_2c_2t_2}{h_1}.
\label{rel65788}
\end{eqnarray}

In general in the dimensional regularization scheme the renormalization constant $Z_3$ can be written as:
\begin{eqnarray}
Z_3=1+\sum_{n=1}^{\infty}\frac{Z_3^{(n)}}{\epsilon^n}.
\label{expr657847566}
\end{eqnarray}
Note that similar expression exist for any renormalization scheme by making simple substitutions for $\epsilon$.

The beta function is defined as:
\begin{eqnarray}
\beta(g^2)=\mu^2\frac{ d  g^2}{d \mu^2}=-\frac{1}{2}g^3\frac{\partial Z_3^{(1)}}{\partial g}=-g^4\frac{\partial Z_3^{(1)}}{\partial  g^2}.
\label{def4456}
\end{eqnarray}

We identify Eq. (\ref{finares4355}) with Eq. (\ref{expr657847566}) noting that the degree of divergence (given by powers in $\frac{1}{\epsilon}$) of the coefficients $u_1$, $v_1$, $v_2$, $v_3$ is zero, one or greater than one:
\begin{eqnarray}
[1+\frac{Z_3^{(1)}}{\epsilon}+\frac{Z_3^{(2)}}{\epsilon^2}+...][1+v_1^{(0)}g^2+\frac{v_1^{(1)}}{\epsilon}g^2+v_2^{(0)}g^4+\frac{v_2^{(1)}}{\epsilon}g^4+
v_3^{(0)}g^6+\frac{v_3^{(1)}}{\epsilon}g^6+...]=u_1^{(0)}g^2+\frac{u_1^{(1)}}{\epsilon}g^2+...
\label{firs5546}
\end{eqnarray}
To order $\frac{1}{\epsilon}$ this leads to:
\begin{eqnarray}
\frac{Z_3^{(1)}}{\epsilon}[1+v_1^{(0)}g^2]+\frac{v_1^{(1)}}{\epsilon}g^2+\frac{v_2^{(1)}}{\epsilon}g^4+\frac{v_3^{(1)}}{\epsilon}g^6=\frac{u_1^{(1)}}{\epsilon}g^2
\label{second6656}
\end{eqnarray}

Here,
\begin{eqnarray}
&&v_1^{(0)}=c_2^{(0)}+t_2^{(0)}+\frac{h_2^{(0)}}{h_1}
\nonumber\\
&&v_3^{(1)}=\frac{(h_2c_2t_2)^{(0)}}{h_1}.
\label{rez43553}
\end{eqnarray}

Using Eq. (\ref{finalrel665466}) and the constraint $c_2=h_4$ by considering simple expansion in powers of $\frac{1}{\epsilon}$ one obtains that $c_2^{(0)}=t_2^{(0)}=h_2^{(0)}=0$.  Eq. (\ref{rez43553}) leads also to $v_3^{(1)}=0$. Then from Eq. (\ref{second6656}) the dependence of $Z_3^{(1)}$ on the coupling constant emerges:
\begin{eqnarray}
Z_3^{(1)}=\beta_0g^2+\beta_1g^4,
\label{dep7768}
\end{eqnarray}
where $\beta_0$ and $\beta_1$ are coefficients independent of $g^2$.
We thus determined the all order shape of the beta function of QCD only by using global properties of the partition function and of the various two, three or four point correlators.

According to Eq. (\ref{def4456}) the beta function contains only the first two order renormalization scheme independent coefficients and is given by:
\begin{eqnarray}
&&\beta(g^2)=\frac{d g^2}{d\ln\mu^2}=
-[\frac{11}{3}N-\frac{2}{3}N_f]\frac{g^4}{16\pi^2}-[\frac{34}{3}N^2-2\frac{N^2-1}{2N}N_f-\frac{10}{3}N N_f]\frac{g^6}{256\pi^4}.
\label{beta435}
\end{eqnarray}

\section*{Acknowledgments} \vskip -.5cm

The work of R. J. was supported by a grant of the Ministry of National Education, CNCS-UEFISCDI, project number PN-II-ID-PCE-2012-4-0078.


\begin{thebibliography}{15}
\bibitem{Vladimirov} A. A. Valdimirov, D. I. Kazakov and O. V. Tarasov, Sov. Phys. JETP {\bf 50} (3), 521 (1979).
\bibitem{Gorishny} S. G. Gorishny, A. L. Kataev, S. A. Larin and L. R. Surguladze, Phys. Lett.  B {\bf 256}, 81 (1991).
\bibitem{Larin1} T. van Ritbergen , J. A. M. Vermaseren and S. A. Larin, Phys. Lett. B {\bf 400}, 379 (1997); arXiv:hep-ph/9701390.
\bibitem{Larin2} J. A. Vermaseren, S. A. Larin and T. van Ritbergen , Phys. Lett. B {\bf 405}, 327-333 (1997); arXiv:hep-ph/9703284.
\bibitem{Kleinert} H. Kleinert, J. Neu, V. Schulte-Frolinde, K. G. Chetyrkin and S. A. Larin, Phys. Lett. B {\bf 272}, 39 (1991); arXiv:hep-th/9503230.
\bibitem{Kataev} A. L. Kataev ans S. A. Larin, JETP Lett. {\bf 96}, 61 (2012); arXiv:1205.2810 [hep-ph].
\bibitem{Baikov} P. A. Baikov, K. G. Chetyrkin, J. H. Kuhn and J. Rittinger, JHEP {\bf 1207}, 017 (2012); arXiv:1206.1284 [hep-ph].

\bibitem{Hooft1} G. 't Hooft, "Some observations in quantum chromodynamics", Notes based on lectures given at Orbis Scientiae 1977, January 17-21, 1977 University of Miami, Coral Gables,
Floride (Reprint of February 1977).
\bibitem{Hooft2} G. 't Hooft, "Can we make sense out of quantum chromodynamics?", Lectures given at Int. School of Subnuclear Physics, Erice, Sisily, July 23-August 10, 1977, PRINT-77-0723(Utrecht) Subnucl. Ser. {\bf 15} (1979) 943.
\bibitem{Jora1} R. Jora and J. Schechter, arXiv:1407.6172[hep-ph] 2014.\
\bibitem{Jora} R. Jora, Int. J. Mod. Phys. A 30, 1550070 (2015), arXiv:1411.0211.
\bibitem{Srednicki} Mark Srednicki, "Quantum Field Theory", Cambridge University Press 2007.
\end{thebibliography}
\end{document}